\documentclass[]{aa}
\usepackage{graphicx,natbib}
\usepackage{amssymb}
\DeclareGraphicsExtensions{.eps,.ps}

\begin{document}
\title{Astrophysical parameters of 14 open clusters projected close to the Galactic plane}

\author{D. Camargo\inst{1} \and C. Bonatto\inst{1} \and E. Bica\inst{1}}

\institute{Universidade Federal do Rio Grande do Sul, Departamento de Astronomia, CP\,15051, RS, Porto Alegre 91501-970, Brazil\\
\email{denilso.camargo@ufrgs.br, \\ bica@if.ufrgs.br, charles@if.ufrgs.br}
\mail{denilso.camargo@ufrgs.br}}

\date{Received --; accepted --}

\abstract
{}
{Astrophysical parameters (\textit{age, reddening, distance, core and cluster radii}) of 14 open 
clusters (OCs) projected close to the Galactic plane are derived with 2MASS photometry. The OCs
are Be\,63, Be\,84, Cz\,6, Cz\,7, Cz\,12, Ru\,141, Ru\,144, Ru\,172, FSR\,101, FSR\,1430, FSR\,1471,
FSR\,162, FSR\,178 and FSR\,198. The OCs Be\,63, Be\,84, Ru\,141, Ru\,144, and Ru\,172 are studied 
in more detail than in previous works, while the others have astrophysical parameters derived for 
the first time.}
{We analyse the colour-magnitude diagrams (CMDs) and stellar radial density profiles (RDPs) built after field-star decontamination and colour-magnitude filtered photometry. Field-star decontamination is applied to uncover the cluster's intrinsic CMD morphology, and colour-magnitude filters are used to isolate stars with a high probability of being cluster members in view of structural analyses.}
{The open clusters of the sample are located at  $d_\odot=1.6-7.1$ kpc from the Sun and at Galactocentric distances $5.5-11.8$ kpc, with age in the range 10 Myr to 1.5 Gyr and reddening 
$E(B-V)$ in the range $0.19-2.56$ mag. The core and cluster radii are in the range $0.27-1.88$\,pc 
and $2.2-11.27$ pc, respectively. Cz\,6 and FSR\,198 are the youngest OCs of this sample, with 
a population of pre-main sequence (PMS) stars, while FSR\,178 is the oldest cluster.}
{}


\titlerunning{Open clusters near the Galactic plane}

\authorrunning{Camargo, Bonatto \& Bica}

\maketitle

%

\section{Introduction}
\label{sec:int}
%

Open clusters (OCs) are self-gravitating stellar systems formed along the gas- and dust-rich Galactic plane. They contain from tens to a few thousand stars distributed in an approximately spherical structure of up to a few parsecs in radius. The structure of most OCs can be roughly described by  two subsystems, the dense core, and the sparse halo \citep[][and references therein]{Bonatto2005}.

Because it is relatively simple to estimate the age and distance of OCs, they have become fundamental probes of Galactic disc properties \citep{Lynga1982, Janes1994, Friel1995, Bonatto2006a, Piskunov2006, Bica2006}. However, the proximity of most OCs to the plane and  the corresponding high values of reddening and field-star contamination  usually restrict this analysis to the more populous and/or 
to those located at most a few kpc from the Sun \citep{Bonatto2006a}.

Detailed analysis of OCs and the derivation of their astrophysical parameters will contribute to future disc studies by unveiling the properties of individual OCs. These parameters, in turn, can help constrain theories of molecular cloud fragmentation, star formation, and dynamical and stellar evolution.

\begin{figure*}
   \centering
   \includegraphics[scale=0.55,viewport=0 0 470 460,clip]{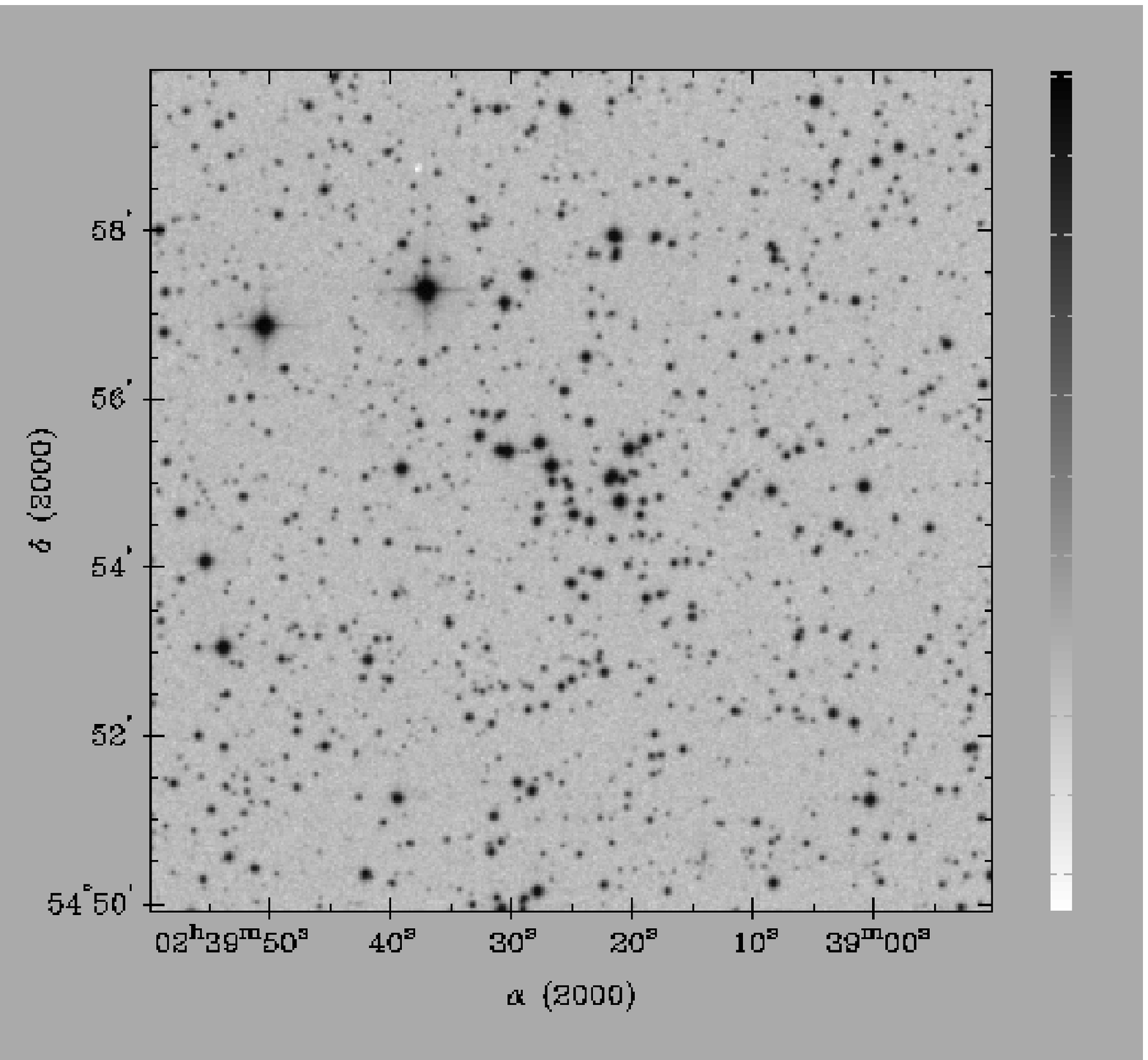}
   \includegraphics[scale=0.55,viewport=0 0 470 460,clip]{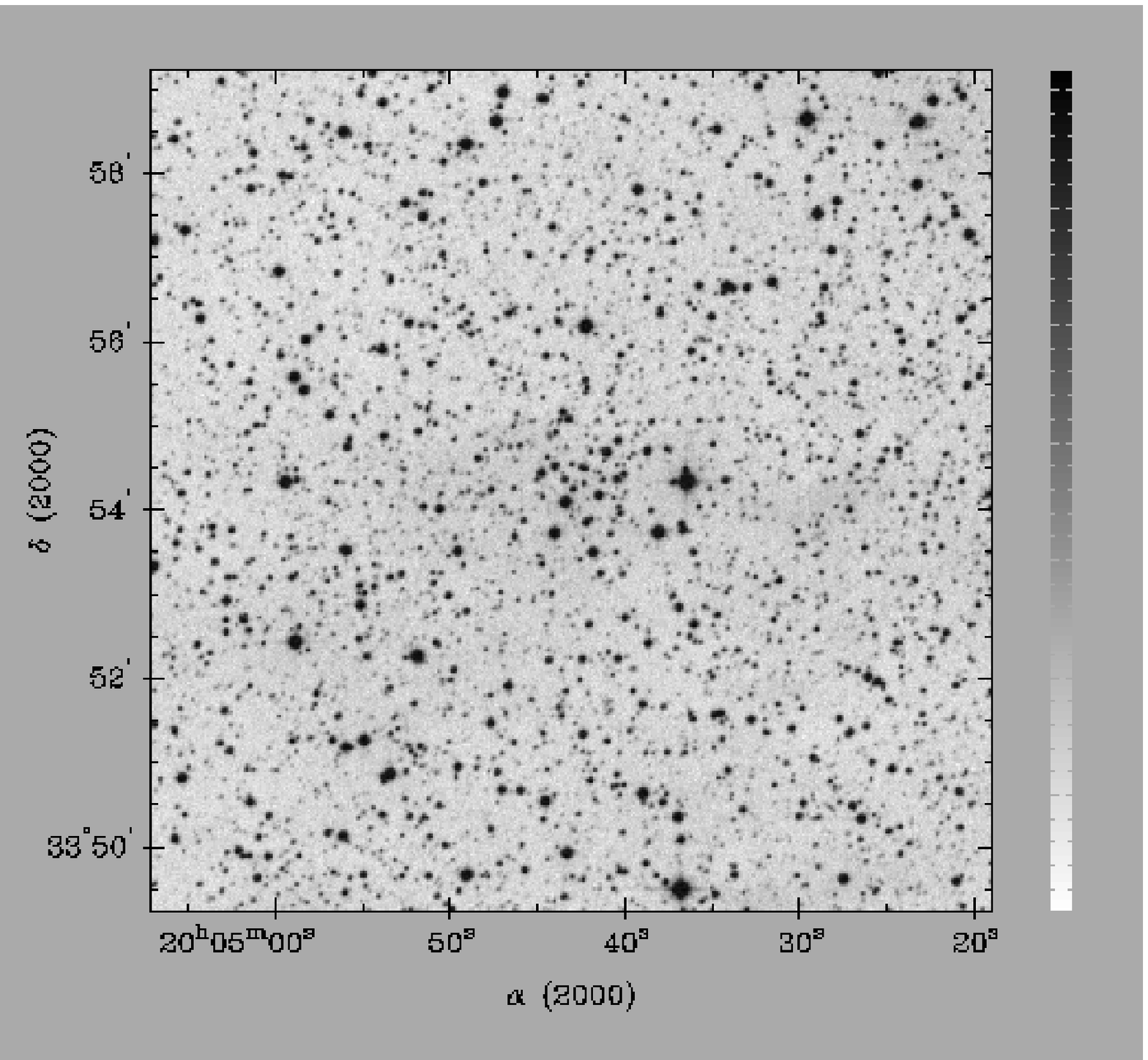}
   \caption[]{Left panel: $10\arcmin\times10\arcmin$ XDSS R image of Cz\,12. Right panel: $10'\times10'$ XDSS R image of Be\,84. Images centred on the optimised coordinates.}
   \label{fig:1}
\end{figure*}

\begin{figure*}
\begin{minipage}[b]{0.50\linewidth}
\includegraphics[width=\textwidth]{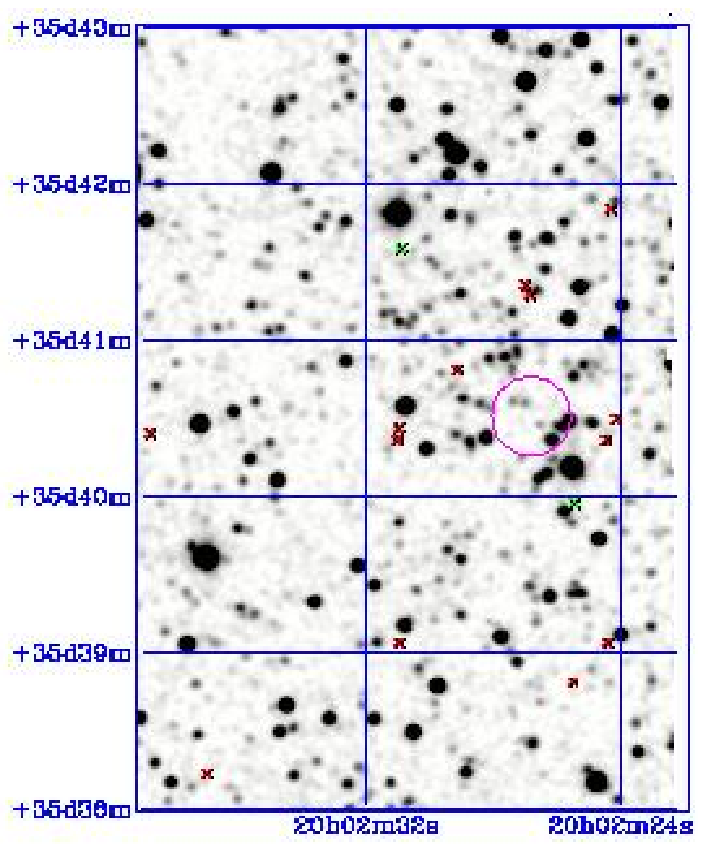}
\end{minipage}\hfill
\begin{minipage}[b]{0.50\linewidth}
\includegraphics[width=\textwidth]{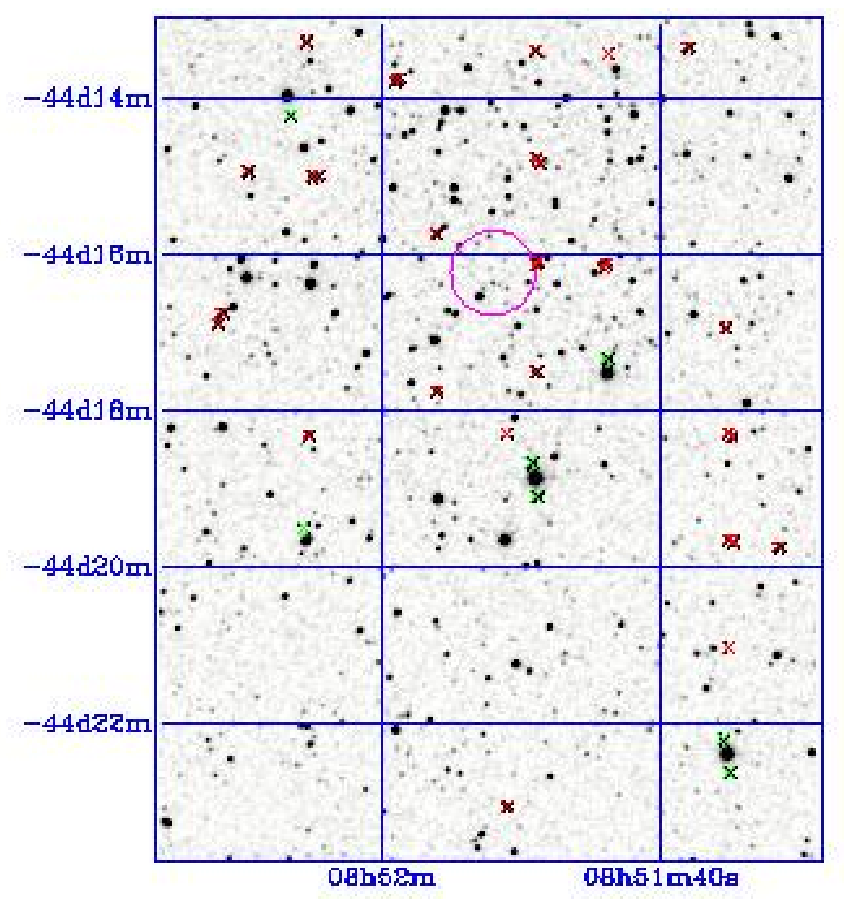}
\end{minipage}\hfill
\caption[]{Left panel: 2MASS K$_S$ image $5\arcmin\times5\arcmin$ of FSR\,198. Right panel: 2MASS 
K$_S$ image $15\arcmin\times15\arcmin$ of FSR\,1430. Images centred on the optimised coordinates. The small circle indicates the cluster central region.}
\label{fig:2}
\end{figure*}

The stellar content of a cluster evolves with time, and internal and external interactions affect the properties of individual clusters. Presently the age distribution of star clusters in the disc of the Galaxy can only be explained if these objects are subjected to disruption timescales of a few times $10^{8}$ yrs \citep{Oort1957, Wielen1971, Wielen1988, Lamers2004}. Open clusters  experience external perturbations by giant molecular clouds (GMCs)  and by spiral arms and other disc-density perturbations. To understand how OCs evolve, it is important to take the effect of these external perturbations into account \citep{Gieles2007}.

\begin{figure}
\resizebox{\hsize}{!}{\includegraphics{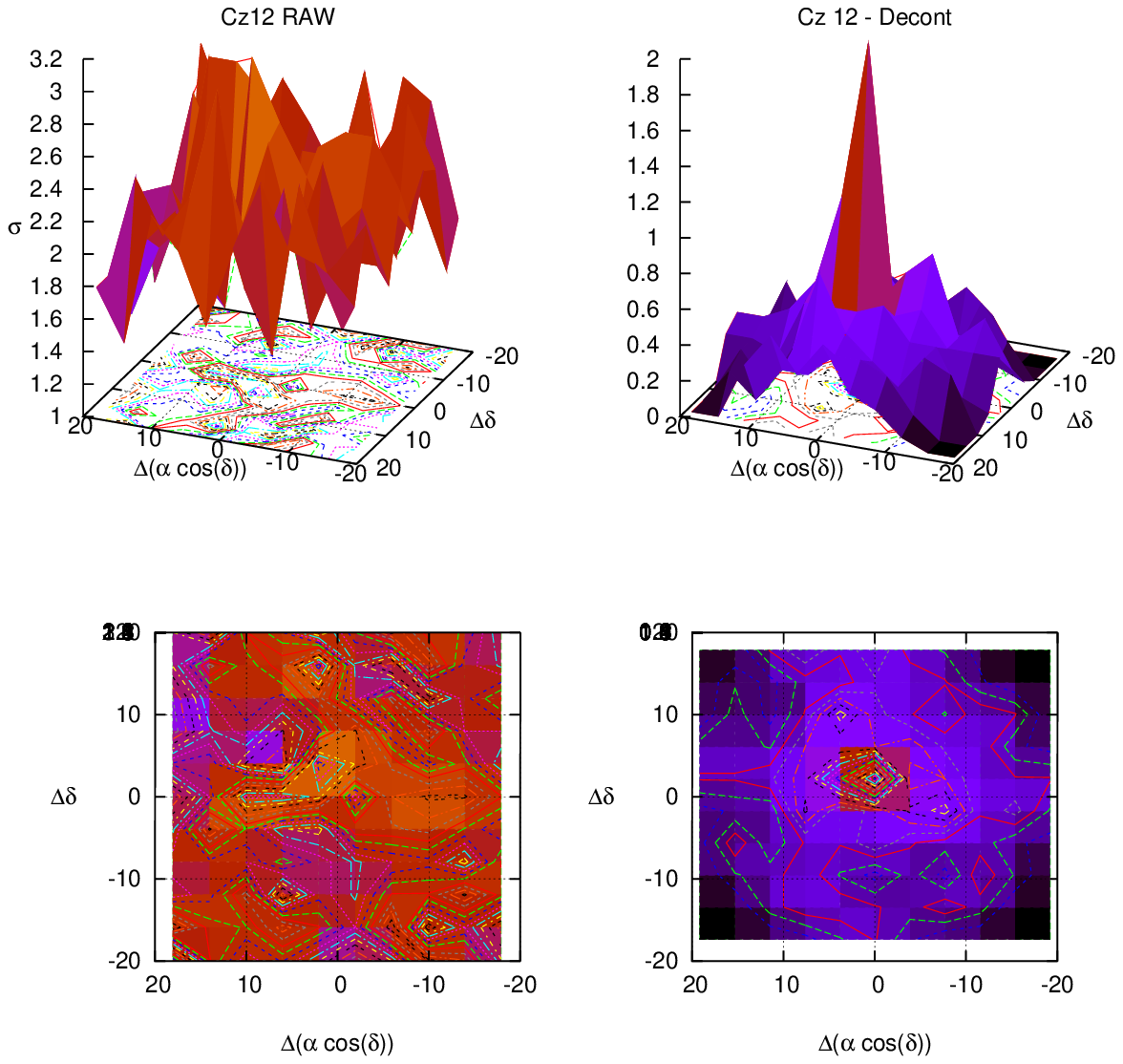}}
   \caption[]{Top panels: stellar surface-density $\sigma (stars\,\rm arcmin^{-2}$) of Cz\,12, computed for a mesh size of $3\arcmin\times3\arcmin$, centred on the coordinates in Table \ref{tab1}. Bottom: the corresponding isopleth surfaces. Left: observed (raw) photometry. Right: Decontaminated photometry.}
   \label{fig:3}
\end{figure}

\begin{figure}
\resizebox{\hsize}{!}{\includegraphics{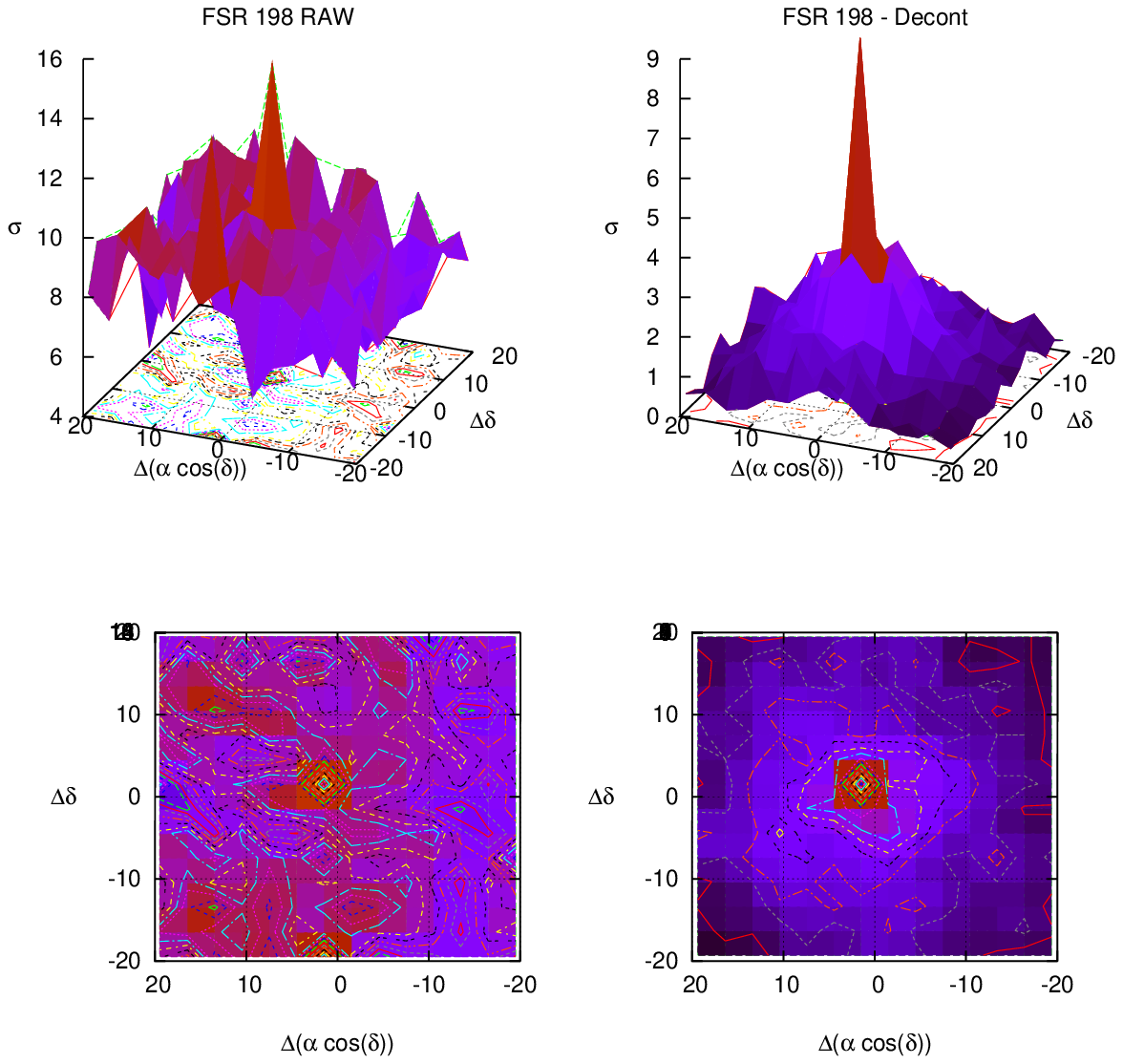}}
   \caption[]{Same as Fig. \ref{fig:3} for  FSR\,198.}
   \label{fig:4}
\end{figure}

Cluster disruption is a gradual process with different mechanisms acting simultaneously. Disruption of OCs due to internal processes are characterised by three distinct phases. These phases and their typical timescales are: (\textit{i}) infant mortality ($\sim10^{7}$ yr), (\textit{ii}) stellar evolution ($\sim10^{8}$ yr) and (\textit{iii}) tidal relaxation ($\sim10^{9}$ yr). During all three phases, there are additional external tidal perturbations from e.g. GMCs and disc-shocking that heat the cluster and speed up the process of disruption. However, these perturbations operate on longer timescales for
cluster populations and so are more important for tidal relaxation \citep{Lamers2004, Lamers2005, Lamers2006}. The combination of these effects results in a time-decreasing cluster mass, until either
its complete disruption or a remnant \citep[][and references therein]{Pavani2007} is left.

\begin{table*}
\centering
{\footnotesize
\caption{Literature and presently optimised coordinates.}
\label{tab1}
\renewcommand{\tabcolsep}{3.6mm}
\renewcommand{\arraystretch}{1.1}
\begin{tabular}{lrrrrrrrrr}
\hline
\hline
\multicolumn{5}{c}{Literature}&\multicolumn{5}{c}{This paper}\\
\cline{2-5}
\cline{7-10}
Cluster&$\alpha(2000)$&$\delta(2000)$&$\ell$&$b$&&$\alpha(2000)$&$\delta(2000)$&$\ell$&$b$\\
&(h\,m\,s)&$(^{\circ}\,^{\prime}\,^{\prime\prime})$&$(^{\circ})$&$(^{\circ})$&&(h\,m\,s)&$(^{\circ}\,^{\prime}\,^{\prime\prime})$&$(^{\circ})$&$(^{\circ})$ \\
\hline
Be\,63 &02 19 36&63 43 00&132.506&2.49& &02 19 30.8&63 43 43&132.49&2.50\\
Be\,84 &20 04 43&33 54 18&70.924&1.27& &20 04 43&33 54 15&70.92&1.27\\
Cz\,6  &02 02 00&62 50 00&130.887&1.05& &02 01 57&62 50 48&130.87&1.07\\
Cz\,7  &02 02 24&62 15 00&131.159&0.52& &02 03 01&62 15 20&131.16&0.53\\
Cz\,12 &02 39 12&54 55 00&138.079&-4.75& &02 39 25&54 54 55&138.11&-4.74\\
Ru\,141 &18 31 19&-12 19 11&19.69&-1.20& &18 31 23&-12 17 50&19.72&-1.21\\
Ru\,144 &18 33 34&-11 25 00&20.749&-1.27& &18 33 33&-11 25 09&20.74&-1.27\\
Ru\,172 &20 11 34&35 35 59&73.11&1.01& &20 11 39&35 37 30&73.14&1.00\\
FSR\,101 &18 49 14&02 46 06&35.147&1.74& &18 49 14&02 46 06&35.14&1.74\\
FSR\,1430 &08 51 52&-44 15 56&264.65&0.08& &08 51 52&-44 16 14&264.66&0.07\\
FSR\,1471 &09 24 08&-47 20 39&270.72&2.14& &09 24 04&-47 20 56&270.71&2.13\\
FSR\,162 &20 01 32&25 14 06&63.211&-2.75& &20 01 26&25 12 30&63.17&-2.74\\
FSR\,178 &20 13 07&29 07 12&67.877&-2.83& &20 13 8.2&29 07 24&67.88&-2.83\\
FSR\,198 &20 02 24&35 41 19&72.184&2.62& &20 02 27&35 40 31&72.17&2.60\\
\hline
\end{tabular}
}
\end{table*}

Probably reflecting the Galactocentric-dependence of most of the disruptive effects, the Galaxy presents a spatial asymmetry in the age distribution of OCs. Indeed, \citet{van1980} noted that OCs older than $\gtrsim1$ Gyr tend to be concentrated in the anti-centre, a region with a low density of GMCs. In this sense, the combined effect of tidal field and encounters with GMCs has been invoked to explain the lack of old OCs in the solar neighbourhood \citep[][and references therein]{Gieles2006}. Near the solar circle most OCs appear to dissolve on a timescale shorter than $\approx1$ Gyr \citep{Bergond2001, Bonatto2006a}. In more central parts, interactions with the disc, the enhanced tidal pull of the Galactic bulge, and the high frequency of collisions with GMCs tend to destroy the poorly populated OCs on a timescale of a few $10^{8}$ yr \citep[e.g.][]{Bergond2001}. \citet{Maciejewski2007} studied a large sample of open clusters, in general not previously studied, to derive fundamental parameters, similar 
to the present analysis.

This paper is organised as follows. In Sect. \ref{sec:targ} we provide general data on the target clusters. In Sect. \ref{sec:tool} we obtain the 2MASS photometry, introduce the tools, CMDs and field-star decontamination algorithm, and derive fundamental parameters of the OCs candidates. In Sect. \ref{sec:stru} we discuss the stellar radial density profiles (RDPs), colour-magnitude filters, and derive structural parameters. In Sect. \ref{rel} we discuss properties of the OCs, and concluding remarks are given in Sect. \ref{conc}.

\section{The target open clusters and candidates}
\label{sec:targ}

The OCs selected for the present analysis are shown in Table \ref{tab1}. These objects are listed in the OC catalogues WEBDA \citep{Mermilliod1996} and \citet{Froebrich2007}. According to the OC catalogues WEBDA \citep{Mermilliod1996} and DAML02 \citep{Dias2002}, the target objects do not have published astrophysical parameters, except for Be\,63, Be\,84, Ru\,141, Ru\,144, and Ru\,172. The optical sample was chosen as a challenge to our analysis tools (Sect.~\ref{sec:tool} and references therein). They are low Galactic-latitude clusters that are often heavily contaminated,  and poorly populated, and that have differential reddening and a few previous parameter determinations if any at all. The infrared candidates (\citealt{Froebrich2007}-FSR) were selected from eye inspections that we made on the 2MASS Atlas for promising candidates. In addition, we analysed some of FSR's quality flag Q0-Q3 objects. All FSR objects that we collected in the present study were concluded to be star clusters (Sects.~\ref{sec:tool} and \ref{sec:stru}).

\citet{Kharchenko2005} employed the ASCC-2.5 catalogue to derive parameters for 520 OCs, using proper motion and photometric criteria to separate probable members from field stars. However, owing to distance and reddening limitations, the fainter cluster parameters rely on a few stars. For Ru\,141 they derived $E(B-V)=0.57$, $d_{\odot}=5.5$ kpc, and age $\thickapprox8$ Myr. For Ru\,172 they derived $E(B-V)=0.20$, $d_{\odot}=1.1$ kpc, and age $\thickapprox0.8$ Gyr.

\citet{Tadross2008} present astrophysical parameters of 24 open clusters of the Berkeley list, using 2MASS photometry and the proper motions of the Naval Observatory Merged Astrometric Dataset (NOMAD). For Be\,63 he derived $E(B-V)=0.90$, $d_{\odot}=3.3$ kpc, $R_{GC}=11.0$ kpc, and age $\thickapprox500$ Myr. For Be\,84 he derived $E(B-V)=0.76$, $d_{\odot}=2.0$ kpc, $R_{GC}=8.1$ kpc, and age $\thickapprox120$ Myr. 

\begin{table}[!h]
\caption{Previous determinations.}
\renewcommand{\tabcolsep}{4mm}
\renewcommand{\arraystretch}{1.1}
\begin{tabular}{lrrrrrr}
\hline
\hline
Cluster&Age&$E(B-V)$&$d_{\odot}$&Source\\
&(Myr)&(mag)&(kpc)\\
($1$)&($2$)&($3$)&($4$)&($5$)\\
\hline
Be\,63 &$500$&$0.90$&$3.3$&(1)\\
Be\,84 &$120$&$0.76$&$2.0$&(1)\\
Ru\,141 &$8$&$0.57$&$5.5$&(2)\\
Ru\,144 &$151$&$0.32$&$-$&---\\
Ru\,172 &$800$&$0.20$&$1.1$&(2)\\
\hline
\end{tabular}
\label{tab2}
\begin{list}{Table Notes.}
\item References: (1) - \citet{Tadross2008}; (2) - \citet{Kharchenko2005}.
\end{list}
\end{table}

In Fig. \ref{fig:1}, we illustrate cluster XDSS\footnote{Extracted from the Canadian Astronomy Data Centre (CADC), at \textit{http://cadcwww.dao.nrc.ca/}} images in the R band of Cz\,12 and Be\,84. In Fig. \ref{fig:2}, we show 2MASS images in the K band of the IR clusters FSR\,198 and FSR\,1430. Be\,84, FSR\,198, and Ru\,144 present significant differential reddening, consistent with their low Galactic latitude (Table \ref{tab1}). Ru\,172 appears to present a similar effect, but only in the background, and the FSR\,1430 image shows a reddening gradient in the north/south direction.
Ru\,144 and Ru\,141 show absorption effects in the background/foreground. Ru\,172 seems to be off-centred in the optical image, probably owing to absorption.

\section{The 2MASS photometry}
\label{sec:tool}
The 2MASS\footnote{The Two Micron All Sky Survey, available at \textit{www..ipac.caltech.edu/2mass/releases/allsky/}} catalogue \citep{Skrutskie2006} was employed in the present work because of the homogeneity and the possibility of large-area data extractions. Also, part of the sample cannot be studied in the optical. VizieR\footnote{http://vizier.u-strasbg.fr/viz-bin/VizieR?-source=II/246.} was used to extract J, H, and ${K_{s}}$ 2MASS photometry. Our previous experience shows that, as long as no other cluster is present in the field and differential absorption is not prohibitive, such large extraction areas provide the required statistics for field-star characterisation.  To maximise the statistical significance and representativeness of background star counts, we use a wide external ring to represent the stellar comparison field. The RDPs produced with the WEBDA coordinates presented in general a dip in the innermost bin. For these we searched for new coordinates that maximise the star-counts at the centre. For each cluster we made circular extractions centred on the optimised coordinates of the clusters. The WEBDA and optimised central coordinates are given in Table \ref{tab1}.

\begin{figure}
\resizebox{\hsize}{!}{\includegraphics{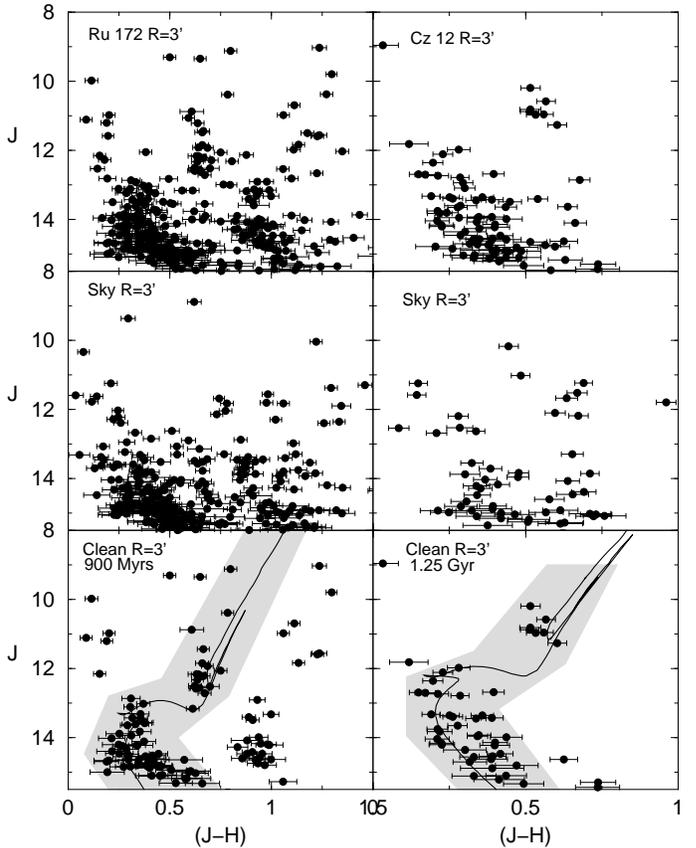}}
   \caption[]{2MASS CMDs extracted from the $R=3'$ region of Ru\,172 and Cz\,12, respectively. Top panels: observed CMDs $J\times(J-H)$. Middle panels: equal area comparison field. Bottom panels: field-star decontaminated CMDs fitted with the 900 Myr Padova isochrone (solid line) for Ru\,172 and 1.25 Gyr for Cz\,12. The colour-magnitude filter used to isolate cluster MS/evolved stars is shown as a shaded region} 
   \label{fig:5}
\end{figure}

\begin{figure}
\resizebox{\hsize}{!}{\includegraphics{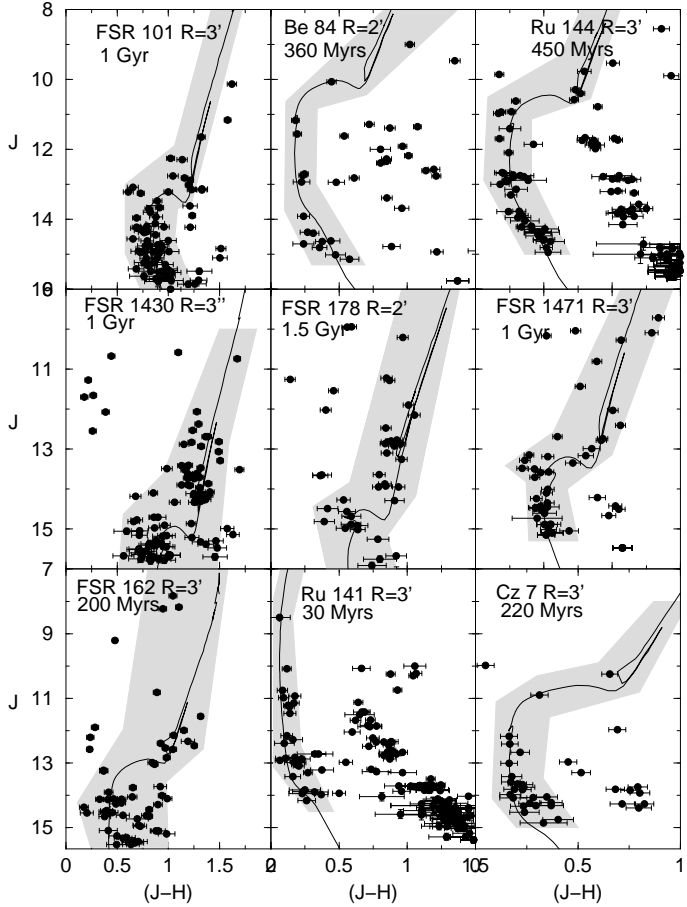}}
   \caption[]{Same as Fig.~\ref{fig:5} for the decontaminated $J\times(J-H)$ CMDs of the central regions of each object.}
   \label{fig:6}
\end{figure}

\begin{figure}
 \resizebox{\hsize}{!}{\includegraphics{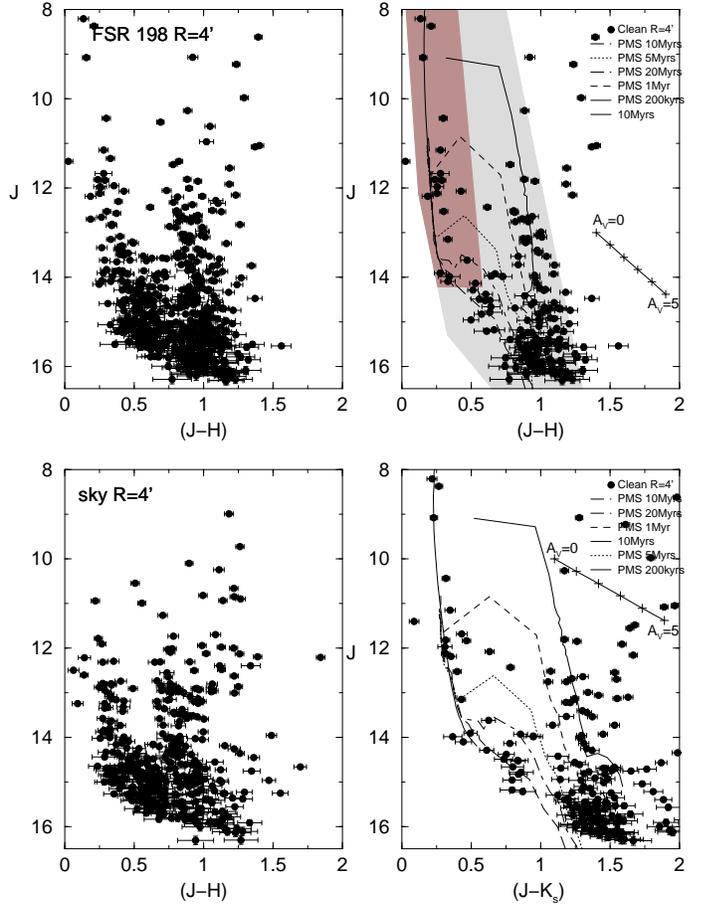}}
   \caption[]{2MASS CMDs extracted from the $R=4'$ region of FSR\,198. Top panels: observed CMDs $J\times(J-H)$ (left) and field star decontaminated CMDs fitted with MS + PMS isochrone solutions. Shaded polygons show the MS (dark) and PMS (light) colour-magnitude filter used to isolate cluster MS/evolved stars (right). Bottom panels: equal area comparison field (left) and $J\times(J-K_{s})$ field star decontaminated CMDs fitted with MS + PMS isochrone solutions.Reddening vectors for $A_V=0-5$ are shown in decontaminated CMDs.}
   \label{fig:7}
\end{figure}

\begin{figure}
 \resizebox{\hsize}{!}{\includegraphics{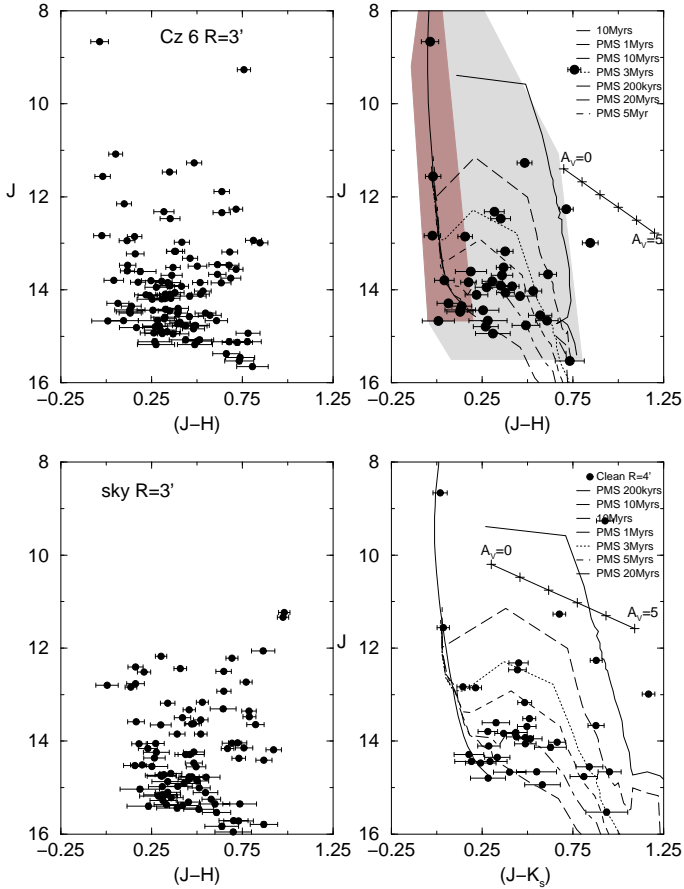}}
   \caption[]{Same as Fig. \ref{fig:7} for CMDs of the central regions ($R=4'$) of Cz\,6. Reddening vectors for $A_V=0-5$ are shown in decontaminated CMDs..}
   \label{fig:8}
\end{figure}

\begin{figure}
 \resizebox{\hsize}{!}{\includegraphics{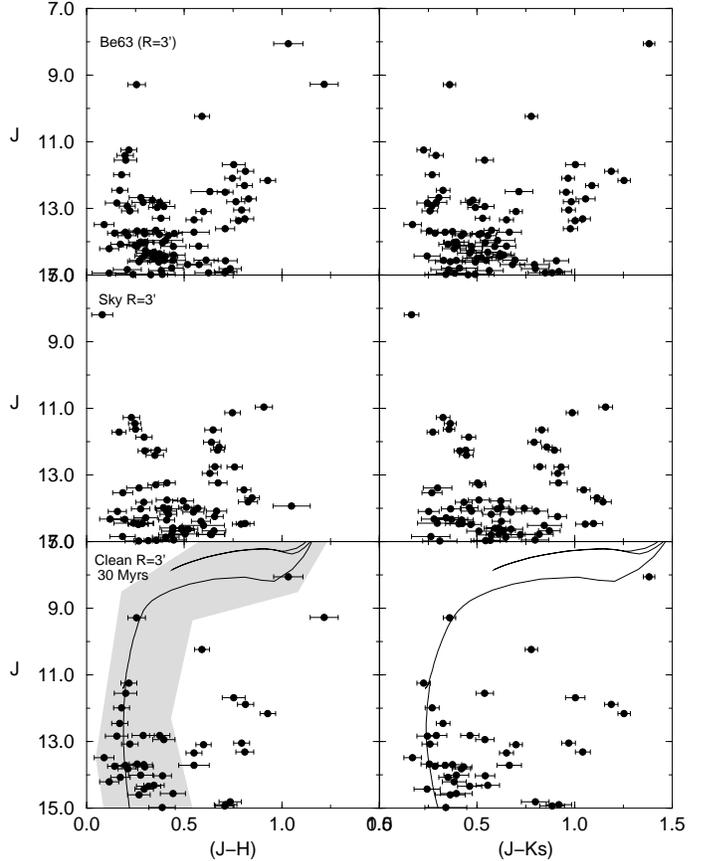}}
   \caption[]{2MASS CMDs extracted from the $R=3'$ region of Be\,63. Top panels: observed CMDs $J\times(J-H)$ (left) and $J\times(J-K_{s})$ (right). Middle panels: equal area comparison field. Bottom panels: field star decontaminated CMDs fitted with the 30 Myrs Padova isochrone (solid line). The colour-magnitude filter used to isolate cluster MS/evolved stars is shown as a shaded region}
   \label{fig:9}
\end{figure}

\begin{figure}
\resizebox{\hsize}{!}{\includegraphics{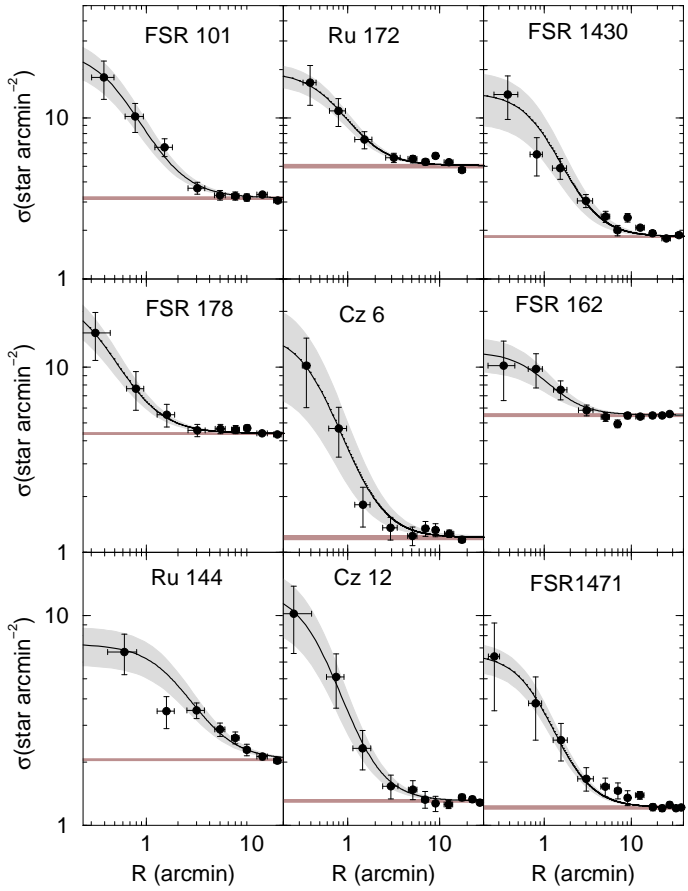}}
   \caption[]{Stellar RDPs (filled circles) built with colour-magnitude filtered photometry. Solid line: best-fit King profile. Horizontal shaded region: stellar background level measured in the comparison field. Gray regions: $1\sigma$ King fit uncertainty.}
   \label{fig:10}
\end{figure}

\begin{figure}
\resizebox{\hsize}{!}{\includegraphics{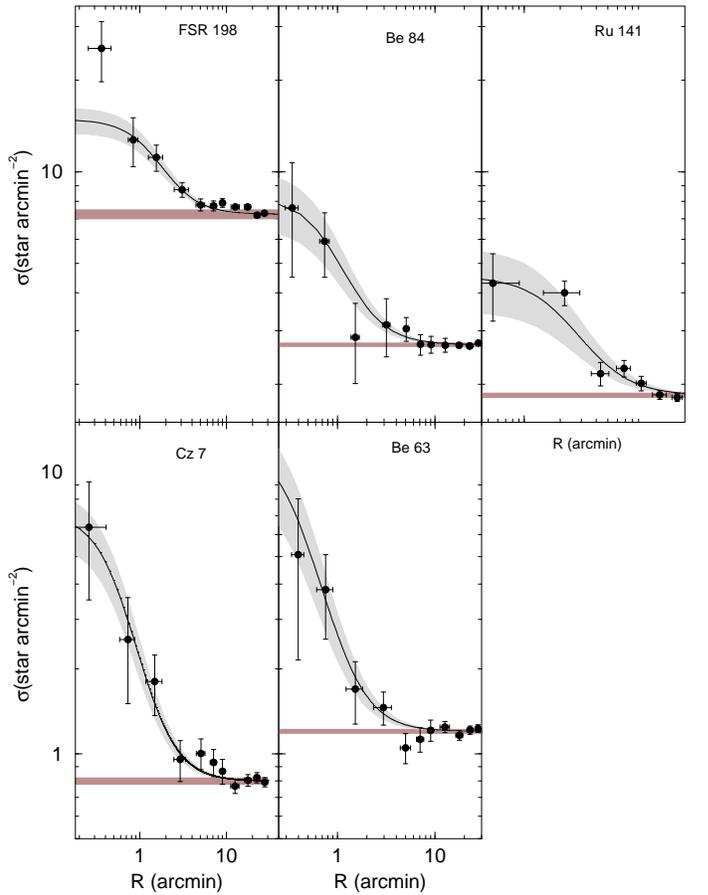}}
   \caption[]{Same as Fig. \ref{fig:10} for the remaining clusters.}
   \label{fig:11}
\end{figure}

\begin{figure}
\resizebox{\hsize}{!}{\includegraphics{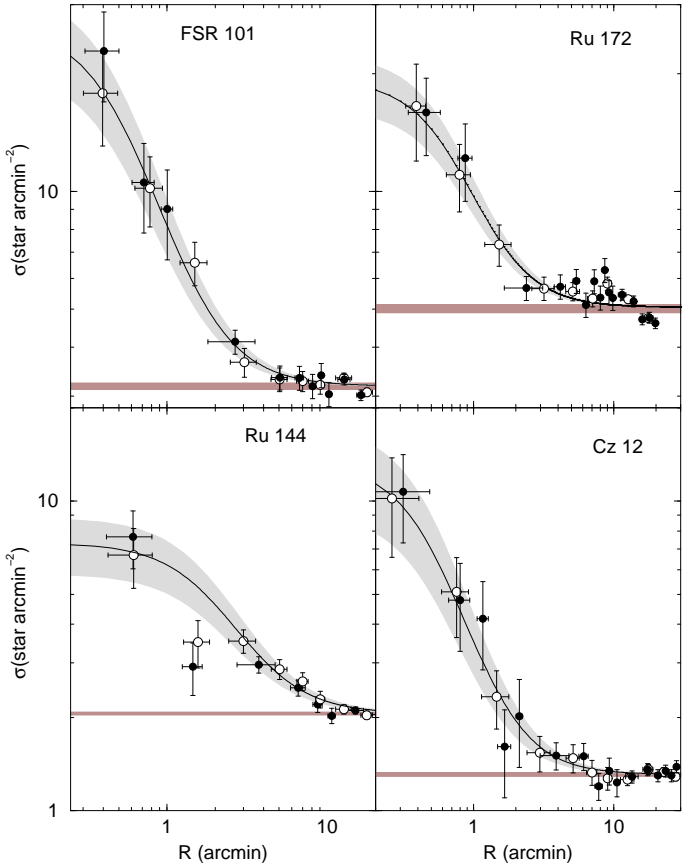}}
   \caption[]{Consistent RDPs are produced with our approach (empty circles) and 
that considering bins containing a fixed number of stars (filled).}
   \label{fig:12}
\end{figure}

\begin{table}[!h]
{\footnotesize
\caption{Derived fundamental parameters.}
\renewcommand{\tabcolsep}{2.0mm}
\renewcommand{\arraystretch}{1.1}
\begin{tabular}{lrrrrr}
\hline
\hline
Cluster&Age&$N_{1\sigma}$&$E(B-V)$&$d_{\odot}$&$R_{GC}$\\
&(Gyr)&&(mag)&(kpc)&(kpc)\\
($1$)&($2$)&($3$)&($4$)&($5$)&($6$)\\
\hline
Be\,63 &$0.03\pm0.01$&3.7&$0.96\pm0.03$&5.7&11.8\\
Be\,84 &$0.36\pm0.05$&5.4&$0.58\pm0.06$&1.7&6.8\\
Cz\,6 &$0.01\pm0.005$&8.3&$0.26\pm0.03$&2.7&6.9\\
Cz\,7 &$0.22\pm0.05$&3.6&$0.70\pm0.03$&3.3&9.7\\
Cz\,12 &$1.25\pm0.4$&4.6&$0.26\pm0.03$&2.0&8.8\\
Ru\,141 &$0.03\pm0.02$&12.8&$0.45\pm0.1$&1.8&5.5\\
Ru\,144 &$0.45\pm0.1$&7.7&$0.77\pm0.1$&1.6&5.7\\
Ru\,172 &$0.9\pm0.2$&5.6&$0.64\pm0.06$&3.1&7.0\\
FSR\,101 &$0.9\pm0.2$&7.1&$2.37\pm0.03$&1.9&7.1\\
FSR\,1430 &$1.0\pm0.3$&8.3&$2.56\pm0.03$&3.6&8.4\\
FSR\,1471 &$1.0\pm0.2$&6.2&$1.22\pm0.02$&2.7&7.7\\
FSR\,162 &$0.2\pm0.05$&3.5&$1.57\pm0.03$&7.1&7.5\\
FSR\,178 &$1.5\pm0.5$&3.9&$1.34\pm0.03$&3.7&6.4\\
FSR\,198 &$0.01\pm0.005$&6.6&$0.96\pm0.03$&1.7&6.9\\
\hline
\end{tabular}
\begin{list}{Table Notes.}
\item The parameter $N_{1\sigma}$  corresponds to the ratio of the number of stars in the decontaminated CMD with respect to the $1\sigma$ Poisson fluctuation measured in the observed CMD \citep{Bica2007}.
Col. 4: reddening in the cluster's central region.
Col. 6: $R_{GC}$ calculated using $R_{\odot}=7.2$ kpc \citep{Bica2006b} as the distance of the Sun to the Galactic centre.
Uncertainties in $d_{\odot}$ and $R_{GC}$ are of the order of $0.1$ kpc.
\end{list}
\label{tab3}
}
\end{table}

The statistical  significance of  astrophysical parameters depends directly on the quality and depth of the photometry \citep{Bonatto2004, Bonatto2005a}. As a photometric quality constraint, 2MASS extractions were restricted to stars with magnitudes (i) brighter than those of the ${99.9\%}$  Point Source Catalogue completeness limit in the cluster direction, and (ii) with errors in J, H, and ${K_{s}}$ smaller than 0.1 mag. The ${99.9\%}$ completeness limits are different for each cluster, varying with Galactic coordinates. A typical distribution of uncertainties as a function of magnitude, for objects projected towards the central parts of the Galaxy, can be found in \citet{Bonatto2007a}. About ${75\%}$ - ${85\%}$ of the stars have errors below 0.06 mag.

\subsection{Field-star decontamination}
\label{sec:red}
The CMD is an important tool for searching for the fundamental parameters of the star clusters, but the field-star contamination is an important source of uncertainty, particularly for low-latitude OCs and/or those projected against the bulge. Our sample of OCs is located in crowded disc zones near the plane, and because of the low latitude, field stars contaminate the CMDs, especially at faint magnitudes and red colours.

To uncover the intrinsic cluster CMD morphology, we use the field-star decontamination procedure described in \citet{Bonatto2007b}, previously applied in the analysis of low-contrast \citep{Bica2005}, embedded \citep{Bonatto2006a}, young \citep{Bonatto2006b}, faint \citep{Bica2006}, old \citep{Bonatto2007a}, or in dense-fields \citep{Bonatto2007b} OCs. The algorithm works on a statistical basis that takes the relative number densities of stars in a cluster region and offset field into account. The algorithm works with three dimensions, the $J$ magnitude and the $(J-H)$ and $(J-K_{s})$ colours, considering as well the respective $1\sigma$ uncertainties in the 2MASS bands. These colours provide the maximum discrimination among CMD sequences for star clusters of different ages \citep[e.g.][]{Bonatto04}.

Basically, the algorithm (i) divides the full range of magnitude and colours of a given CMD into a 3D grid whose cubic cells have axes along the $J$, $(J-H)$, and ${(J-K_{s})}$ directions, (ii) computes the expected number density of field stars in each cell based on the number of comparison field stars (within  $1\sigma$ Poisson fluctuation) with magnitude and colours compatible with those of the cell, and (iii) subtracts the expected number of field stars from each cell. Consequently, this method is sensitive to local variations in field star contamination with magnitude and colours. Cell dimensions are $\Delta{J}=1.0$, and  $\Delta(J-H)={\Delta(J-K_{s})}=0.15$, which are adequate to allow sufficient star-count statistics in individual cells and preserve the morphology of the CMD evolutionary sequences. The dimensions of the colour/magnitude cells can be changed subsequently so that the total number of stars subtracted throughout the whole cluster area matches the expected one, within the $1\sigma$ Poisson fluctuation.

Three different grid specifications in each dimension are used to minimise potential artifacts introduced by the choice of parameters, thus resulting in 27 different outputs. They occur because for a CMD grid beginning at magnitude $J_0$ (with cell width $\Delta\,J$), we also include additional runs for cell centres shifted by $J_0\pm\frac{1}{3}\Delta\,J$. Also when considering the same strategy applied to the
2 colours, we end up with 27 outputs. The average number of probable cluster stars $\langle{N_{cl}}\rangle$ is computed from these outputs. Typical standard deviations of $\langle{N_{cl}}\rangle$ are at the $\approx2.5\%$ level. The final field-star decontaminated CMD contains the $\langle{N_{cl}}\rangle$ stars with the highest number frequencies. Stars that remain in the CMD after the field star decontamination are in cells where the stellar density presents a clear excess over the field. Consequently, they have a significant probability of being cluster members. Further details on the algorithm, including discussions of subtraction efficiency and limitations, are given in \citet{Bonatto2007b}.

\citet{Bica2007} introduce the parameter $N_{1\sigma}$ which corresponds to the ratio of the number of stars in the decontaminated CMD with respect to the $1\sigma$ Poisson fluctuation measured in the observed CMD. By definition, CMDs of overdensities must have $N_{1\sigma}>1$. It is expected that CMDs of star clusters have $N_{1\sigma}$ significantly larger than 1. The $N_{1\sigma}$ values for the present sample are given in col. 3 of Table \ref{tab2}.

\subsection{Fundamental parameters}
\label{sec:fun}
Astrophysical fundamental parameters are derived with solar-metallicity Padova isochrones \citep{Girardi2002} computed with the 2MASS $J$, $H$, and $K_{s}$ filters. The 2MASS  transmission filters produced isochrones very similar to the Johnson-Kron-Cousins ones \citep[e.g.][]{Bessel1988}, with differences of at most 0.01 in $(J-H)$ \citep{Bonatto04}. The best fits are superimposed on decontaminated CMDs.  Parameters derived from the isochrone fit are the observed distance modulus $(m-M)_{J}$ and reddening $E(J-H)$, which converts to $E(B-V)$ and $A_{V}$ with the relations $A_{J}/{A_{V}}=0.276$, $A_{H}/{A_{V}}=0.176$, $A_{K_{s}}/{A_{V}}=0.118$, $A_{J}=2.76\times{E(J-H)}$, and $E(J-H)=0.33\times{E(B-V)}$ \citep{Dutra2002}, assuming a constant total-to-selective absorption ratio $R_{V}=3.1$.
The resulting age, $E(B-V)$, $d_{\odot}$, and $R_{GC}$ are given in cols. 3 to 6 of Table 2. FSR\,198 and Cz\,6 presents a significant population of pre-main sequence (PMS) stars. Isochrones of \citet{Siess2000} are used to characterise the PMS sequences of these objects (Figs.~\ref{fig:7} and \ref{fig:8}).

\begin{figure}
\resizebox{\hsize}{!}{\includegraphics{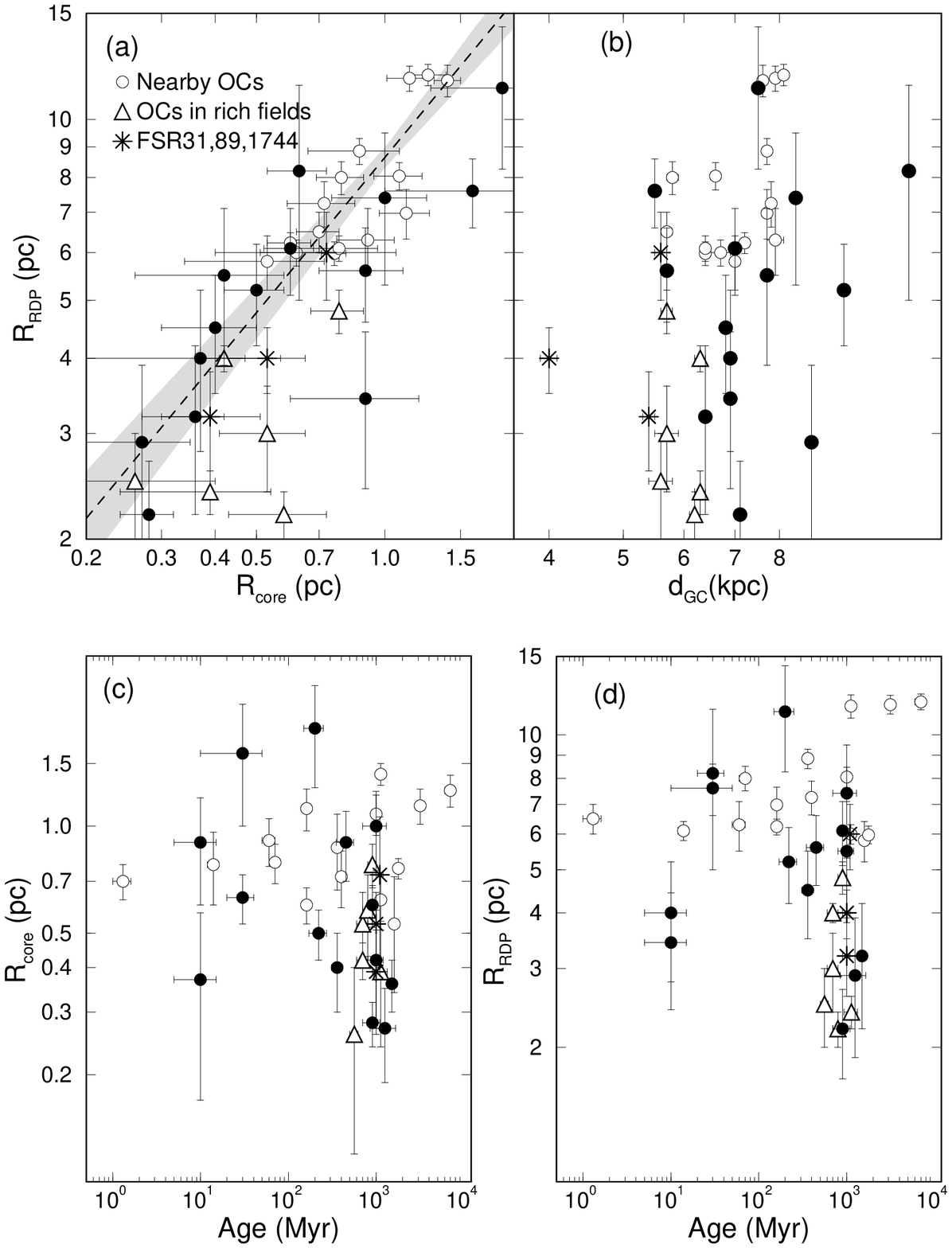}}
\caption[]{Relations involving structural parameters of OCs. Empty circles: nearby OCs, including two young ones. Triangles: OCs projected on dense fields towards the centre. Stars: the similar OCs FSR\,31, FSR\,89 and FSR\,1744. Black circles: the present work OCs.}
\label{fig:13}
\end{figure}

In Fig. \ref{fig:5} we present the $J\times(J-H)$ CMDs extracted from a region $R=3'$ centred on the optimised coordinates of Ru\,172 and Cz\,12 (top-panel). In the middle panels we show the background field corresponding to a ring with the same area as the central region. In the bottom panels we built the field-star decontaminated CMDs.

\begin{table*}
{\footnotesize
\begin{center}
\caption{Structural parameters.}
\renewcommand{\tabcolsep}{1.1mm}
\renewcommand{\arraystretch}{1.1}
\begin{tabular}{lrrrrrrrrrrr}
\hline
\hline
Cluster&$(1')$&$\sigma_{0K}$&$\sigma_{bg}$&$R_{core}$&$R_{RDP}$&$\sigma_{0K}$&$\sigma_{bg}$&$R_{core}$&$R_{RDP}$&${\Delta}R$&CC\\
&($pc$)&($*\,pc^{-2}$)&($*\,pc^{-2}$)&($pc$)&($pc$)&($*\,\arcmin^{-2}$)&($*\,\arcmin^{-2}$)&($\arcmin$)&($\arcmin$)&($\arcmin$)&\\
($1$)&($2$)&($3$)&($4$)&($5$)&($6$)&($7$)&($8$)&($9$)&($10$)&($11$)&($12$)\\
\hline
Be\,63&$1.64$&$4.2\pm1.5$&$0.4\pm0.01$&$0.6\pm0.1$&$8.2\pm3.2$&$11.5\pm4.0$&$1.2\pm0.02$&$0.38\pm0.08$&$5.0\pm2.0$&$15-30$&0.92\\
Be\,84 &$0.50$&$21.9\pm6.9$&$10.6\pm0.1$&$0.4\pm0.1$&$4.5\pm1.0$&$5.5\pm1.7$&$2.6\pm0.02$&$0.2\pm0.1$&$2.2\pm1.0$&$15-30$&0.89\\
Cz\,6&$0.81$&$21.8\pm12.0$&$1.85\pm0.05$&$0.37\pm0.2$&$4.0\pm1.2$&$14.2\pm7.8$&$1.2\pm0.03$&$0.46\pm0.2$&$5.0\pm1.5$&$10-20$&0.90\\
Cz\,7&$0.95$&$6.8\pm1.6$&$0.9\pm0.02$&$0.5\pm0.08$&$5.2\pm1.0$&$6.28\pm1.5$&$0.8\pm0.02$&$0.53\pm0.09$&$5.5\pm2.0$&$10-30$&0.95\\
Cz\,12&$0.57$&$35.8\pm12.6$&$3.98\pm0.06$&$0.27\pm0.08$&$2.9\pm1.0$&$11.8\pm4.13$&$1.31\pm0.02$&$0.5\pm0.2$&$5.0\pm2.0$&$10-20$&0.88\\
Ru\,141&$0.53$&$2.6\pm1.4$&$0.9\pm0.1$&$1.6\pm0.6$&$7.6\pm1.0$&$0.7\pm0.4$&$0.25\pm0.02$&$3.0\pm1.2$&$14.5\pm2.0$&$10-20$&0.85\\
Ru\,144 &$0.47$&$23.8\pm6.8$&$9.0\pm0.4$&$0.9\pm0.2$&$5.6\pm1.0$&$5.3\pm1.5$&$2.0\pm0.08$&$1.95\pm0.36$&$12.0\pm2.0$&$12-20$&0.97\\
Ru\,172 &$0.89$&$17.7\pm3.7$&$6.3\pm0.06$&$0.6\pm0.08$&$6.1\pm1.0$&$14.2\pm3.0$&$6.3\pm0.06$&$0.7\pm0.1$&$6.8\pm1.0$&$8-20$&0.96\\
FSR\,101 &$0.54$&$79.7\pm21.3$&$10.8\pm0.4$&$0.28\pm0.04$&$2.2\pm0.5$&$23.5\pm6.3\pm$&$3.2\pm0.1$&$0.52\pm0.09$&$4.0\pm1.0$&$8-20$&0.97\\
FSR\,1430 &$1.05$&$11.9\pm5.0$&$1.5\pm0.05$&$1.0\pm0.25$&$7.4\pm2.1$&$13.0\pm5.4$&$1.7\pm0.05$&$0.95\pm0.24$&$7.0\pm2.0$&$20-40$&0.87\\
FSR\,1471 &$0.78$&$9.9\pm4.95$&$1.6\pm0.05$&$0.42\pm0.16$&$5.5\pm1.6$&$6.1\pm3.0$&$0.97\pm0.03$&$0.53\pm0.2$&$7.0\pm2.0$&$20-40$&0.80\\
FSR\,162 &$2.05$&$1.55\pm0.6$&$1.3\pm0.01$&$1.88\pm0.6$&$11.27\pm3.0$&$6.54\pm2.5$&$5.5\pm0.05$&$0.92\pm0.3$&$5.5\pm1.5$&$15-30$&0.88\\
FSR\,178 &$1.07$&$17.9\pm5.0$&$3.8\pm0.05$&$0.36\pm0.06$&$3.2\pm1.0$&$20.6\pm5.7$&$4.4\pm0.06$&$0.34\pm0.06$&$3.0\pm0.5$&$10-20$&0.97\\
FSR\,198 &$0.49$&$30.7\pm5.6$&$29.2\pm0.2$&$0.7\pm0.1$&$3.43\pm1.0$&$7.6\pm1.4$&$7.2\pm0.06$&$1.5\pm0.2$&$7.0\pm2$&$15-30$&0.98\\
\hline
\end{tabular}
\begin{list}{Table Notes.}
\item Col. 2: arcmin to parsec scale. To minimise degrees of freedom in RDP fits with the King-like profile (see text), $\sigma_{bg}$ was kept fixed (measured in the respective comparison fields) while $\sigma_{0}$ and $R_{core}$ were allowed to vary. Col. 11: comparison field ring. Col. 12: correlation coefficient.
\end{list}
\label{tab4}
\end{center}
}
\end{table*}

Both Ru\,172 and Cz\,12 can be recognised as a cluster by the presence of the MS and a prominent giant clump. These features are not present in the comparison field (Fig. \ref{fig:5} middle panels). Figure~\ref{fig:6} shows 9 OCs. FSR\,162 and FSR\,178 are probable remnant OCs. The present sample of OCs has in general low stellar-density contrast with respect to the background owing to the projection close to the plane (Table \ref{tab1}). 

Cz\,7 is not a populous cluster, but the results point to a relatively young OC (age $\approx220$ Myrs). The decontamination leads to an age of $\approx30$ Myrs for Be\,63, but deep observations are required for more conclusive results. Deeper photometry is essential in most cases, especially for faint and/or distant OCs, close to the plane, affected by 2MASS completeness limits. Cz\,6 and FSR\,198 are very young  OCs (10 Myrs). Be\,84 turns out to be moderately young (age $\approx360$ Myr), similar to Ru\,144 with 450 Myrs. After decontamination, Ru\,141 corresponds to the bluest sequence in that diagram (age $\approx30$ Myr). Since Ru\,141, Ru\,144, and Ru\,172 are located at very low Galactic latitudes (Table~\ref{tab1}), important absorption variations across the cluster area and/or background may occur, which can produce residual effects in decontaminated CMDs. In particular, Ru\,141 has a strong absorption in J for the background stars at $\approx2.5\arcmin$ to the northeast. It also has a strong absorption in B at $\approx6\arcmin$ to the east. Indeed, Ru\,141 shows significant residuals in the CMD (Fig. \ref{fig:6}). We point out that the field of Ru\,141 also contains the OC Ru\,142 at $\approx$ $10\arcmin$. Finally, FSR\,162 is a faint and distant OC ($d_{\odot}=7.1$kpc). Only field decontamination made possible to probe cluster properties. Their OC nature is further supported by their decontaminated structural properties (Sect.~\ref{sec:stru}). 

We note that there are some differences in the fundamental parameters with respect to 
previous works (e.g. \citealt{Kharchenko2005}; \citealt{Tadross2002}), especially  
for cluster age. This occurs especially for young clusters in which field contamination
has not been properly taken into account. In these cases, PMS stars in conjunction with 
important field contamination may mimic older ages. A clear example is FSR\,198
(Fig.~\ref{fig:7}). 

\section{Structural parameters}
\label{sec:stru}
Structural parameters have been derived by means of the stellar RDPs, defined as the projected number of stars per area around the cluster centre. RDPs are built with stars selected after applying the respective CM filter to the observed photometry. The colour-magnitude filters (CM filters) are shown in Figs. \ref{fig:5} - \ref{fig:9} as the shaded region superimposed on the field-star decontaminated CMDs.

Colour-magnitude filters are only used to discard stars with colours comparable to those of the foreground/background field. This tool was previously applied to the structural analysis of the OCs M67 \citep{Bonatto2003}, NGC\,3680 \citep{Bonatto2004}, NGC\,188 \citep{Bonatto2005a}, NGC\,6611 \citep{Bonatto2006}, NGC\,4755 \citep{Bonatto2006b}, M52 and NGC\,3690 \citep{Bonatto2006c} and the faint OCs BH 63, Lyng\aa \, 2, Lyng\aa \, 12 and King 20 \citep{Bica2006}. The filters were defined based on the distribution of the decontaminated star sequences in the CMDs of open clusters. They are wide enough to accommodate cluster MS and evolved star colour distributions, allowing for $1\sigma$ photometric uncertainties. CM filter widths should also account for formation or dynamical evolution-related effects, such as enhanced fractions of binaries (and other multiple systems) towards the central parts of clusters, since such systems tend to widen the MS \citep[e.g.][]{Bonatto2007b, Bonatto2005a, Hurley1998, Kerber2002}. 
However, residual field stars with colours similar to those of the cluster are expected to remain inside the CM filter region. They affect the intrinsic RDP to a degree that depends on the relative densities of field and cluster stars. The contribution of the residual contamination to the observed RDP is statistically considered by means of the comparison field. In practical terms, the use of CM filters in cluster sequences enhances the contrast of the RDP against the background level, especially for objects in dense fields \citep[see e.g. Sect. \ref{sec:stru} in ][]{Bonatto2007b}.

\begin{figure}
\resizebox{\hsize}{!}{\includegraphics{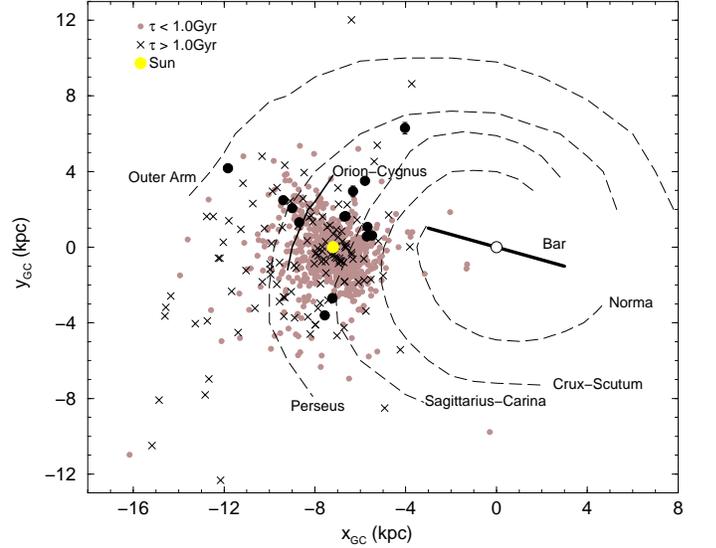}}
\caption[]{Spatial distribution of the present star clusters (filled circles) compared to the WEBDA OCs with ages younger than 1 Gyr (gray circles) and older than 1 Gyr (`x'). The schematic projection of the Galaxy is seen from the North pole, with 7.2 kpc as the Sun's distance to the Galactic centre.}
\label{fig:14}
\end{figure}

To avoid oversampling near the centre and undersampling for large radii, the RDPs were built by counting stars in concentric rings of increasing width with distance to the centre. The number and width of rings are adjusted so that the resulting RDPs present adequate spatial resolution with moderate $1\sigma$ Poisson errors. The $R$ coordinate (and
respective uncertainty) of a given ring corresponds to the average distance to the
cluster centre (and standard deviation) computed for the stars within the ring. The residual background level of each RDP corresponds to the average number of CM-filtered stars measured in the comparison field. Alternatively, we build RDPs with bins of variable sizes to
check for any systematic biases that may have been introduced by our method.
Following \citet{MA05} and \citet{MK09}, we computed the RDPs with bins that contain a fixed 
number of stars, 10 for $0<R(\arcmin)<1$, 100 for $1<R(\arcmin)<10$, and 1000 for 
$R>10\arcmin$. Within the uncertainties, both approaches produce similar RDPs, as shown
by the examples illustrated in Fig.~\ref{fig:12}.

Structural parameters were derived by fitting the two-parameter \citet{King1966a} surface-density profile to the colour-magnitude filtered RDPs. The two-parameter King model essentially describes the intermediate and central regions of globular clusters \citep{King1966b, Trager1995}. 
The fit was performed using a nonlinear least-squares fit routine that uses the errors as weights. The best-fit solutions are shown in Figs. \ref{fig:10} and \ref{fig:11} as a solid line superimposed on the RDPs. King's law is expressed as $\sigma(R)=\sigma_{bg}+\sigma_{0K}/(1+(R/R_{core})^{2}$, where $\sigma_{bg}$ is the background surface density of stars, $\sigma_{0K}$ is the central density of stars and $R_{core}$ is the core radius. The cluster radius ($R_{RDP}$) and uncertainty can be estimated by considering the fluctuations of the RDPs with respect to the residual background, and $R_{RDP}$ corresponds to the distance from the cluster centre where RDP and comparison field become statistically indistinguishable. The structural parameters derived are given in Table \ref{tab4}. 

FSR\,198 presents a conspicuous excess over the King-like profile in the innermost RDP bin (Fig. \ref{fig:11}). Such a feature was been attributed to advanced dynamical evolution, having been detected in post-core collapse globular clusters \citep{Trager1995}. Some Gyr-old OCs, such as NGC3960 \citep{Bonatto2006c} and LK 10 \citep{Bonatto2009a}, also present this cusp. However, very young OCs such as NGC\,2244 \citep{Bonatto2009b}, NGC\,6823 \citep{Bica2008}, Pismis\,5, and NGC\,1931 \citep{Bonatto2009c} also display the RDP excess. Consequently, molecular cloud fragmentation and/or star formation effects probably play an important role in shaping the early stellar radial distribution of some OCs.

\section{Relations among astrophysical parameters}
\label{rel}

At this point it is interesting to compare the structural parameters derived for the present OCs with those measured in different environments (Fig. \ref{fig:13}). We considered (\textit{i}) a sample of bright nearby OCs \citep{Bonatto2005}, including the two young OCs NGC\,6611 \citep{Bonatto2006}, and NGC\,4755 \citep{Bonatto2006b}, (\textit{ii}) OCs projected against the central parts of the Galaxy \citep{Bonatto2007b}, and (\textit{iii}) the recently analysed OCs FSR\,1744, FSR\,89 and FSR\,31 \citep{Bonatto2007a} projected against the central parts of the Galaxy, and (\textit{iv}) the present sample.

The comparison OCs in the sample (\textit{i}) have ages in the range $70\,Myr$ - $7\,Gyr$, masses within $400-5300M_{\odot}$, and Galactocentric distances in the range $5.8\la{R_{GC}}(kpc)\la8.1$. NGC\,6611 has $\thickapprox1.3$ Myr and $R_{GC}=5.5$ kpc, and NGC\,4755 has $\thickapprox14$ Myr and $R_{GC}=6.4$ kpc. Sample (\textit{ii}) OCs are characterised by $600\,Myr\la{age}\la1.3\,Gyr$ and $5.6\la{R_{GC}}(kpc)\la6.3$. Sample (\textit{iii}) consists of Gyr-class OCs at $4.0\la{R_{GC}}(kpc)\la5.6$. 

In panel (\textit{a}) of Fig. \ref{fig:13}, core and cluster radii of the OCs in sample (\textit{i}) are almost linearly related by $R_{RDP}=(8.9\pm0.3)\times R_{core}^{(1.0\pm0.1)}$, which suggests that both kinds of radii undergo a similar scaling, in the sense that on average, larger clusters tend to have larger cores. However, $\frac{1}{3}$ of the OCs in sample (\textit{ii}) do not follow that relation, which suggests that they are either intrinsically small or have been suffering important evaporation effects. The core and cluster radii in sample (\textit{iii}) and the OCs of this work (\textit{iv}) are consistent with the relation  at the $1\sigma$ level. A dependence of OC size on Galactocentric distance is shown in panel (\textit{b}), as previously suggested by \citet{Lynga1982} and \citet{Tadross2002}. In panels (\textit{c}) and (\textit{d}) we compare core and cluster radii with cluster age, respectively. This relationship is intimately related to cluster survival/dissociation rates. Both kinds of radii present a similar dependence on age, in which part of the clusters expand with time, while some seem to shrink. The bifurcation occurs at an age $\approx1$ Gyr. A similar effect was observed for the core 
radii of LMC and SMC star clusters (e.g. \citealt{MG03}), which have core radii ($\rm0.5\la R_c(pc)\la8$) and 
mass ($\rm10^3\la M(M_\odot)\la10^6$) significantly more than the present ones. The core radii distribution of 
most LMC and SMC clusters is characterised by a trend toward increasing core radius with age with an apparent
bifurcation (core shrinkage) at several hundred Myr. \citet{MG03} argue that this relationship represents 
true physical evolution, with some clusters developing expanded cores due to the stellar mass
black-holes, and some that contract because of dynamical relaxation and core collapse (\citealt{MG08}). 
We also note that the radii of the young clusters (age $<20$\,Myr) of our sample are related to
the age similarly to the {\em leaky} ones of \citet{Pfl09}. Similar relations involving core and 
cluster radii were found by \citet{Maciejewski2007} for an optical cluster sample. 


Finally, Fig. \ref{fig:14} shows the spatial distribution in the Galactic plane of the present OCs, compared to that of the OCs in the WEBDA database. We consider two age ranges, $<1$ Gyr and $>1$ Gyr. We compute the projections on the Galactic plane of the Galactic coordinates $(\ell,b)$. Old OCs are primarily found outside the solar circle, and the inner Galaxy contains the few OCs detected so far. The interesting point here is whether inner Galaxy clusters cannot be observed because of strong absorption and crowding, or have been systematically dissolved by the different tidal effects combined \citep[][ and references therein]{Bonatto2007a}. In this context, the more OCs identified (with their astrophysical parameters derived) in the central parts, the more constraints can be established to settle this issue.

Differential reddening provides uncertainties in OC astrophysical parameters. Most OCs of our sample occur close to spiral arms. Since they are located close to the plane (Table. \ref{tab1}), they may have interacted with the arms, especially by means of encounters with GMCs. 

\section{Concluding remarks}
\label{conc}
In the present work, we have derived astrophysical parameters of 14 OCs projected close to the Galactic plane by means of 2MASS CMDs and stellar RDPs. Field-star decontamination is applied to uncover the cluster's intrinsic CMD morphology, and CM filters are used to isolate probable cluster members. That field star decontamination leads to consistent CMDs and RDPs shows that we are dealing with OC, instead of field fluctuations. In particular, the present CMD and RDP analyses indicate that 6 IR objects from \citet{Froebrich2007}, initially identified as cluster candidates from overdensities, are star clusters (Table \ref{tab4}). 

Our sample contains OCs with ages in the range $10\pm5$ Myr (Cz\,6 and FSR\,198) to $1.5\pm0.5$ Gyr (FSR\,178), at distances from the Sun in the range $d_{\odot}\thickapprox1.6$ kpc (Ru\,144) to $d_{\odot}\thickapprox7.1$ kpc (FSR\,162) and Galactocentric distances $R_{GC}\thickapprox5.5$ kpc for Ru\,141 to $R_{GC}\thickapprox11.8$ kpc for Be\,63.

Be\,84, Ru\,141, and Ru\,144 are relatively young surviving OCs located inside the solar circle. Clusters in that region are expected to suffer important tidal stress in the form of shocks from disc and bulge crossings, as well as encounters with massive molecular clouds. In the long run, these processes tend to dynamically heat a star cluster, which enhances the rate of low-mass star evaporation and produces a cluster expansion on all scales. However, for some clusters, mass segregation and evaporation may also lead to a phase of core contraction. Consequently, these effects tend to disrupt most clusters, especially the less populous ones. On the other hand, FSR\,1430, FSR\,1471 and Cz\,12 are older. One of the reasons for such longevity may be their large Galactocentric distance, which minimises the disruption effects. The newly formed open clusters Cz\,6 and FSR\,198 show PMS stars in the CMDs.

The present study contributes new open cluster parameters and some revisions to the DAML02 and WEBDA open cluster databases.
\vspace{0.8cm}

\textit{Acknowledgements}: 
We thank an anonymous referee for suggestions.
We acknowledge support from the CNPq and CAPES (Brazil).

\end{document}